# A Microscopy Approach to Investigating the Energetics of Small Supported Metal Clusters

*Barbara A. J. Lechner, Fabian Knoller, Alexander Bourgund, Ueli Heiz, and Friedrich Esch\**

Chair of Physical Chemistry, Department of Chemistry & Catalysis Research Center, Technical University of Munich, Lichtenbergstr. 4, 85748 Garching, Germany

\* friedrich.esch@tum.de

ABSTRACT. Metal clusters are partway between molecular and bulk systems and thus exhibit special physical and chemical properties. Atoms can rearrange within a cluster to form different structural isomers. Internal degrees of freedom and the interaction with the support – which both are dependent on cluster size – can promote diffusion across a support. Here, we show how fast scanning tunneling microscopy (FastSTM) can be used to investigate such dynamical behavior of individual clusters on the example of $Pd_n$ (1≤n≤19) on a hexagonal boron nitride nanomesh on Rh(111), in particular pertaining to minority species and rare events. By tuning the cluster size and varying the temperature to match the dynamics to the FastSTM frame rate, we followed steady state diffusion of clusters and atoms inside the nanomesh pore and reversible cluster isomerization *in situ*. While atoms diffuse along the rim of a pore, a small cluster experiences a corrugation in the potential energy landscape and jumps between six sites around the center of the pore. The atom and cluster diffusion between pores is strongly influenced by defects.

INTRODUCTION

Small metal clusters have discrete electronic states like molecules and as a result, very different properties from their bulk counterparts: Au becomes a highly active and selective catalyst,[1–3] Pt becomes non-metallic,[4] and Ag and Au can appear in different colors, covering the entire visible range[5]. In optical devices and heterogeneous catalysis, their superior characteristics have long made clusters and nanoparticles indispensable. A key challenge to working with clusters lies in controlling their size and shape down to a single atom. Yet, even when the cluster has been created and deposited onto a suitable support material, it can still change its shape by isomerization or its size by sintering, which can lead to deactivation of a catalyst.[6] The two main pathways for sintering are the detachment of atoms from one cluster and attachment to another, called Ostwald ripening,[7] and the diffusion of entire clusters to form larger agglomerates, called Smoluchowski ripening.[8] Furthermore, clusters have been shown to exhibit a structural fluxionality in the presence of reactants leading to low temperature catalysis,[9,10] an idea which has recently gained renewed interest.[11] To control the cluster shape and prevent sintering, it is of paramount importance to gain a fundamental understanding of the energetics driving cluster dynamics. In that respect, several interactions are of relevance, including the cluster–support interaction strength, energy dissipation from a cluster into the support, the distribution of energy into the internal vibrational modes of a cluster and the response of a cluster to an electrostatic field. By observing processes such as isomerization between different configurations of the same cluster or diffusion on the support, we can infer the driving forces for a process. Therefore, the aim of the present work is to observe the dynamics of different size clusters microscopically.

Scanning tunneling microscopy (STM) has been used successfully to monitor surface dynamics for many years: the diffusion of molecules[12] and metal adatoms[13] has been observed in time-lapse images, the rotational motion of molecules induced and monitored[14] and mobile species followed directly by the STM tip.[15,16] Clusters have been investigated in several STM studies as well,[17] ranging from the formation of clusters from individual atoms[18] to the observation of sintering of size-selected, deposited clusters.[19,20] In the present study, we use a standard Omicron instrument which we have equipped with an additional electronics module to operate the instrument at elevated scan speeds. In our FastSTM measurements, we reach close to video frame rates, which enables us to observe processes in situ that could not be resolved with standard scan speeds and to dynamics which only occur at higher temperatures.

For a first explorative FastSTM investigation of cluster dynamics, we deposited Pd clusters on a hexagonal boron nitride (h-BN) film on Rh(111) which decreases the bond strength of the cluster to the underlying metal support and introduces a periodic wettability that is high in the center of the pores and low on the rim. Due to the different lattice constant of BN and Rh, the film forms a moiré network with a corrugation of approx. 0.055 nm and a periodicity of 3.2 nm. This so-called nanomesh was first described by Corso and coworkers[21] and its atomic model later unraveled by Laskowski and coworkers.[22,23] Its unit cell consists of 13 x 13 BN units on 12 x 12 Rh atoms and contains a hexagonal depression of 2.0 nm in diameter where the N atoms are strongly bound to the underlying Rh – termed "pore" –, surrounded by a less strongly bound, elevated edge – termed "wire". The h-BN film can be synthesized to a high quality, with long-range order and minimal defect density. However, looking at STM images in the literature, point-defects appearing as depressions on a three-fold corner of the wires and domain boundaries are still relatively common. A very precise measurement of the defect density was reported by Mertens who used Cu underpotential deposition on highly-ordered h-BN films on Rh(111) to estimate a defect fraction on the order of 0.08–0.7% of the geometric area,[24] i.e. up to one defect per moiré unit cell on average. Theoretical studies investigating the nucleation and stability of defects in boron nitride have reported different types of defects in the nanomesh pores.[25] Small amounts of C contaminants can create point defects:[26,27] a B vacancy – which does not cost much, energetically[28] – can be filled by a C atom. Such '$C_B$' defects have been shown to be energetically stable and catalytically active sites by density functional theory (DFT) calculations and appear preferentially at the rim of a pore.[26] BN vacancy pairs are other energetically stable defects, again located preferentially at the rim.[28] These sites expose the underlying Rh substrate and are thus good candidates for binding Pd clusters strongly. We will see below how such defects can influence the dynamics of Pd adatoms and clusters.

Due to the periodic arrangement of pores, the nanomesh has been widely used as a substrate in recent years, *e.g.* serving as a template for nanoparticle growth.[29] Here, defects are less important since the sample is heated and sufficient metal deposited to fill each pore with a similar amount of material. Molecular adsorbates preferably sit on the rim of a pore rather than at its center, and adsorption on the elevated wires is generally avoided.[30–32] Dil and coworkers explained this phenomenon with a different work function on wires and pores, leading to a gradient in the electrostatic potential at the rim of each pore.[33] Xe atoms and phthalocyanine molecules, which are only weakly physisorbed, thus reside preferentially at the location with the largest electrostatic potential gradient. In the present paper, the nanomesh pores serve as microscopic cells in which we can contain $Pd_n$ clusters while keeping their interaction with the metal substrate low to study the dynamics of the cluster and their related energetics. First, we look at the size-dependent stability of the clusters and recap the known sintering behavior. We then look at cluster dynamics which we tune to the frame rates of our FastSTM measurements by matching temperature and cluster size – atoms are investigated at cryogenic temperatures while the sample is heated to induce the diffusion of a cluster. In this manner, we observe *in situ* a range of cluster dynamics, from isomerization events to steady state diffusion of clusters and atoms.

METHODS

The h-BN nanomesh was synthesized by exposing a freshly cleaned Rh(111) single crystal surface to $6 \times 10^{-7}$ mbar borazine, $(BH)_3(NH)_3$, at 1113 K for 3 min. Cleaning of the Rh(111) crystal was achieved by 10 min sputtering in $5 \times 10^{-6}$ mbar Ar at 1.5 kV, subsequent temperature cycling between 773 K and 1253 K in $1 \times 10^{-7}$ mbar $O_2$, and a final annealing step to 1253 K in ultra-high vacuum (UHV). Size-selected Pd clusters were produced in a high-frequency laser ablation source.[34] Thereby, the second harmonic of a Nd:YAG laser is focused onto a rotating Pd target and the resulting plasma cooled in an adiabatic expansion of He (grade 6.0) to enable cluster creation and form a directional beam. The clusters are then focused by electrostatic lenses, selected for positive charges by a 90° electrostatic quadrupole bender and mass-selected in a quadrupole mass filter, before being soft-landed on the h-BN film (<1 eV/atom). The clusters were deposited at 150 K for low temperature STM measurements and at room temperature (RT) for measurements at RT and above. Cluster coverages were kept low to prevent a statistically significant number of clusters being deposited in the same nanomesh pore.

All STM measurements were performed with an Omicron VT-AFM instrument using an etched Pt/Ir tip (Unisoku). Sample cooling was achieved with a continuous flow of liquid $N_2$, sample heating using a boron nitride heater located underneath the Rh(111) crystal. A type K thermocouple connected to the sample was used to measure temperatures above RT, a diode on the sample cooling plate for cryogenic temperatures. According to Omicron specifications for the relation between sample temperature and diode reading, we estimate that our low temperature (LT) measurements were at approx. 150 K. The addition of a FAST-module[35] to the standard Omicron controller allowed us to drive the STM at elevated scan rates (several frames per second). In a FastSTM movie, the fast scan direction is driven in a sinusoidal pendulum motion and the feedback slowed down to correct only for drift and creep but not follow the topography of the surface, resulting in quasi-constant height measurements. In order to interact with a cluster in a rather controlled way, we lowered the loop gain of the feedback (*i.e.* slowed it down) *before* starting the FastSTM measurement. In this way, the tip approaches the cluster apex considerably while the pendulum motion starts, facilitating atom transfer from the cluster to the tip. We used this method to shrink a cluster *in situ* and thus make it more mobile, as shown in Figures S1 and S2. In standard STM mode, constant current measurements were performed. These details and the tunneling parameters are given in each figure caption. Frame rates of time series measurements and FastSTM movies are given in frames per minute (fpm) and frames per second (fps), respectively.

Image processing was performed in Gwyddion,[36] using plane subtraction, facet leveling and row-by-row alignment for background correction and the affine distortion and drift correction functions to remove drift from LT images. We determined the height distribution of clusters by detecting particles based on an intensity threshold, drawing a profile that cuts through the cluster maximum and determining the peak height with respect to the background around the cluster. To account for the difference of the determined background level (on the rim of the nanomesh) and the actual minimum where the clusters sit (in the pores) as well as for electronic effects, 0.13 nm were added to the measured values to obtain the actual cluster height.[20] FastSTM movies were recorded as a one-dimensional data stream and reconstructed into an image stack, background leveled and drift corrected with a specially written python software package.

RESULTS AND DISCUSSION

**a) Size-dependent cluster stability**

Deposition of differently sized $Pd_n$ clusters onto h-BN/Rh(111) shows that the size-selection is maintained on the surface for clusters between two and 19 atoms,[20] while single atoms ($Pd_1$) undergo significant sintering and grow into clusters of varying size at RT. To judge cluster size from an STM image, we use the apparent height, or brightness, since the apparent lateral extension results from a convolution of tip and cluster and is therefore strongly dependent on tip state. Figure 1 shows Pd clusters covering the h-BN film randomly across several terraces of the underlying Rh(111) crystal. The clusters in Figure 1b-h exhibit visibly different sizes, their brightness ranging from just above the corrugation of the nanomesh for 1- to 2-atom species to more than two substrate steps for 19-atom clusters. This observation is quantified in Figure 1i where the height distributions are presented in histograms. The approximate height for single- and multilayer clusters is marked by red dashed lines.[20] Ignoring the topmost histogram, the average cluster height increases gradually from $Pd_1$ to $Pd_{19}$ and ranges from one to three layers. The width of each histogram distribution spans at most two different apparent sizes, which are due to structural isomers, *e.g.* two and three layer clusters for $Pd_9$ and $Pd_{12}$ and three and four layer clusters for $Pd_{19}$. The small number of clusters higher than 1.3 nm for $Pd_{19}$ are consistent with the Poisson probability of depositing two or more clusters in the same nanomesh pore at the relatively high coverage employed here. Single-layer species appear for $Pd_1$ to $Pd_4$, yet they do not all have the same apparent height: While $Pd_1$ and $Pd_2$ just protrude from the nanomesh pores, resulting in height distributions with a peak below the one-layer mark, the onset of the height histograms for $Pd_3$ and $Pd_4$ is shifted to larger heights. This difference in contrast can be explained by a lower density of states of the smaller species.

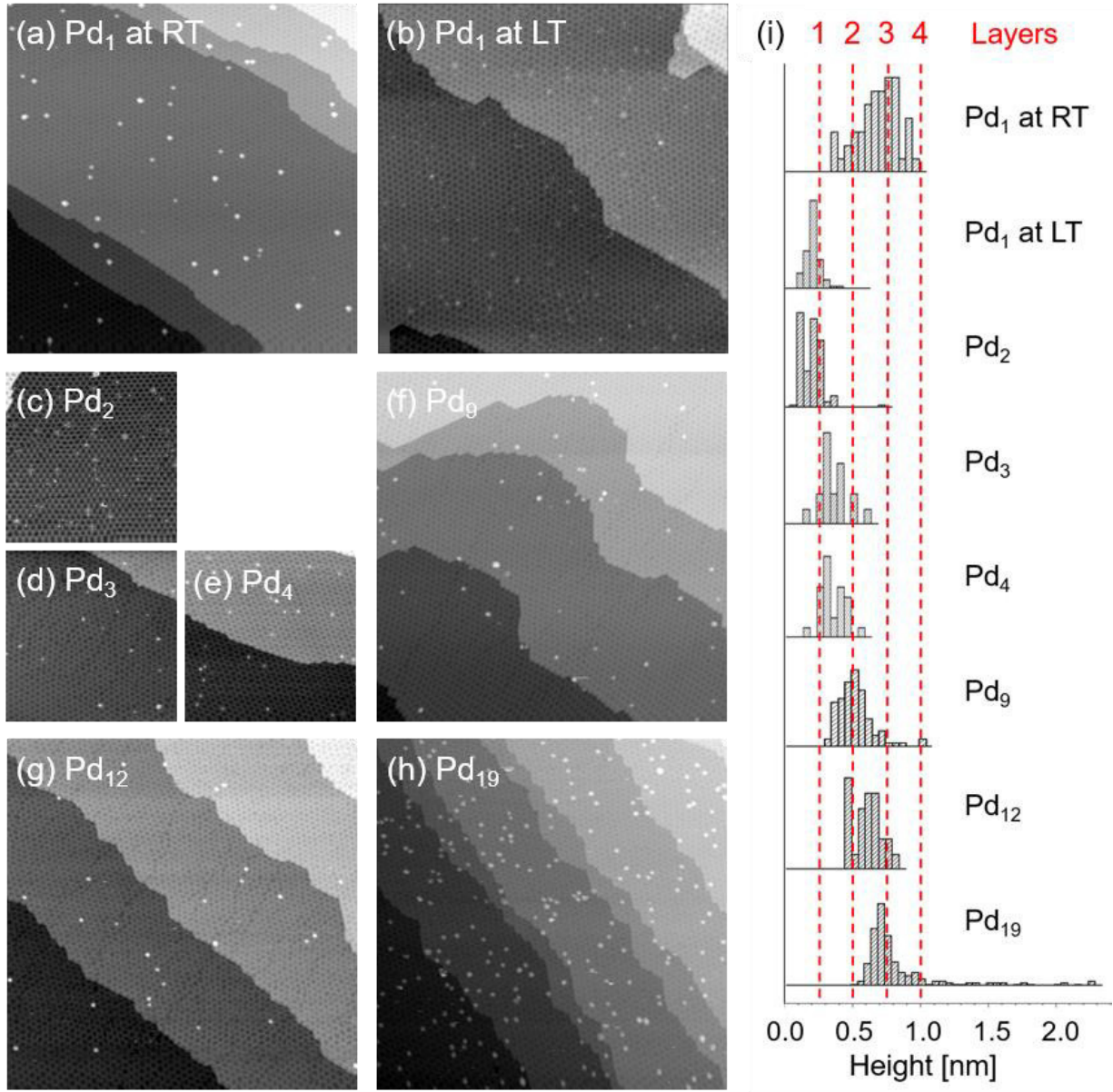

Figure 1. The appearance of different size Pd clusters on h-BN/Rh(111) in STM. (a) $Pd_1$ at RT, (b) $Pd_1$ at LT, (c) $Pd_2$ at LT, (d) $Pd_3$ at RT, (e) $Pd_4$ at RT, (f) $Pd_9$ at RT, (g) $Pd_{12}$ at RT, (h) $Pd_{19}$ at RT. (i) Height profiles comparing the size distribution of clusters in (a-h). *Imaging parameters: $V_b$ = -0.6 V; (a-c, e, f, h) $I_t$ = 1 pA, (d) $I_t$ = 50 pA, (g) $I_t$ = 150 pA; (a, b, f-h) 200 x 200 $nm^2$, (c-e) 100 x 100 $nm^2$.*

Atoms deposited at RT (Figure 1a and top histogram in Figure 1i) result in a distribution of cluster sizes up to four layers. The atoms have sintered significantly. This observation is in good agreement with previous work by our group which showed that Pd clusters exhibit exclusively Ostwald ripening, *i.e.* diffusion of single atoms to increase the size of larger clusters while dissolving smaller ones, on h-BN/Rh(111),[20] and with DFT calculations which showed that Pd adatoms can diffuse between pores at RT.[37] The present data shows that the rate-limiting step in the Ostwald ripening process is the detachment of atoms from clusters – hindered at RT – while the diffusion of atoms occurs facilely. Diffusion and consequently sintering of Pd atoms can be suppressed by cooling the h-BN sample to approx. 150 K during deposition and STM measurements, as shown in Figure 1b. Upon heating the sample to RT, the atoms sinter into clusters with a distribution comparable to that of $Pd_1$ deposited at RT. For direct comparison of atoms and clusters, we also studied $Pd_2$ at the same low temperature (Figure 1c) and confirmed that they have a similar overall appearance and comparable height distribution. The facile diffusion of atoms compared to small clusters can be explained not only by the changing interaction (or footprint) but also by the distribution of energy into their degrees of freedom (DOF), where atoms only exhibit translational DOF while clusters also have additional internal DOF. As a result, it is more likely for sufficient thermal energy to migrate into a translational, *i.e.* directional DOF in the case of an atom than in the case of a cluster, resulting in a higher probability of diffusion for atoms.

b) **Dynamics of Pd atoms**

We now have a look at the dynamics of the most mobile Pd species on h-BN/Rh(111): the Pd atom. Analysis of the distribution of $Pd_1$ at LT with a total coverage of 10% of the nanomesh unit cells (see Figure 1b and Figure S3) reveals that a surprisingly large number of pores seemingly accommodates multiple atoms. In principle, the appearance of double and triple protrusions could be an STM image of a single atom switching between two positions faster than the imaging rate. However, the concerted flipping of the species marked '2' in Figure 2a between two locations in a pore (yellow arrow) makes this very unlikely since a single atom would need to move on two vastly different time scales to explain this behavior. We therefore propose that each protrusion is in fact a Pd atom. Employing this model, we count 6.1% singly, 2.3% doubly and 0.5% triply occupied pores, which is a significantly higher multiple occupancy than predicted by Poisson statistics for randomly deposited atoms (9.5%, 0.5% and 0.02%, respectively). In addition, ~1.4% of pores are occupied by species of unclear shape (*e.g.* rings, which we will discuss below). It therefore appears that the atoms do not remain where they landed during deposition but instead diffuse across the surface until they are immobilized. Defects such as BN vacancy pairs and $C_B$ defects, which we discussed in the Introduction, are good candidates to immobilizing Pd atoms at the rim of a pore. Two atoms are often located close to each other within a pore, which could be explained by a defect appearing more likely next to another one. Interestingly, it is preferable for two atoms to reside next to each other rather than fuse into a dimer (which would appear as single protrusions, as seen in Figure 1c). Hence, there must be an energy barrier which prevents dimer formation and thus cluster nucleation at cryogenic temperatures. This suppression of the tendency to coalesce is an indication for the strong influence of defects on the Pd atom diffusion on h-BN in general. Indeed, several of our images reveal a local variation in pore size and shape (*e.g.* Figures 2 and S3) which is consistent with the presence of defects in the h-BN film.[28]

In addition to the point-like protrusions, we observe a small number of rings for $Pd_1$/h-BN. One such example is labelled '4' in Figure 2a. Similar ring shaped species in STM images of adsorbates on h-BN/Rh(111) have been reported previously. Mn, Co and Fe exhibited two states on h-BN, a ring state and a single protrusion.[31] These two states could be reversibly converted into each other by voltage pulses and current ramps. The authors concluded that the transition metal adatom could alter the bond strength of the h-BN film to the substrate and thus lead to detachment of the film, resulting in an electronic state imaged as a ring. In another STM study, Xe atoms deposited onto h-BN/Rh(111) at 5 K were found to adsorb preferably at the rim of the nanomesh pores.[30] While static at 5 K, these atoms became mobile at somewhat higher temperatures and thus appeared as a continuous ring. To determine whether the Pd ring species in the present case are electronic or dynamic in nature, we turned to FastSTM and imaged them with a higher time resolution. Figure 2b shows that a ring which appeared as a full circle – if a bit fuzzy – in standard STM (labelled 'std.'), indeed appears only as a partial ring in faster recorded frames. Moreover, the part of the circle which is bright changes from frame to frame, which is a clear indication that the ring is a fast-moving species running around the rim of the pore it occupies. The apparent average diameter of ring species is $2.0 \pm 0.3$ nm which matches the pore diameter of 2.2 nm well.[32] Even though we still do not measure fast enough to resolve the motion of the atom fully, it shows clearly that the ring state in our case is a dynamic adatom. By averaging all frames of Supporting Movie 1, we obtain a map of the likelihood with which the cluster resides in different parts of the pore – higher brightness means longer overall residence in that location. The thus obtained ring labelled 'ave.' shows a picture very similar to the standard STM image, and reveals two locations with low residence probability. We conclude that the pore is not perfectly symmetric and that this distortion causes a local preference of some adsorption sites over others. Accordingly, we can understand partial rings as adatoms exploring defective pores. In contrast to the FastSTM frames of Figure 2b, the partial ring in Figure 2c (left) is localized in the same part of the pore in consecutive frames (not shown) and the pore visibly blocked where the ring ends. It seems as though a mobile atom remains mobile for longer than the average time it needs to meet a defect site, implying that there is an inward barrier to binding to the defect.

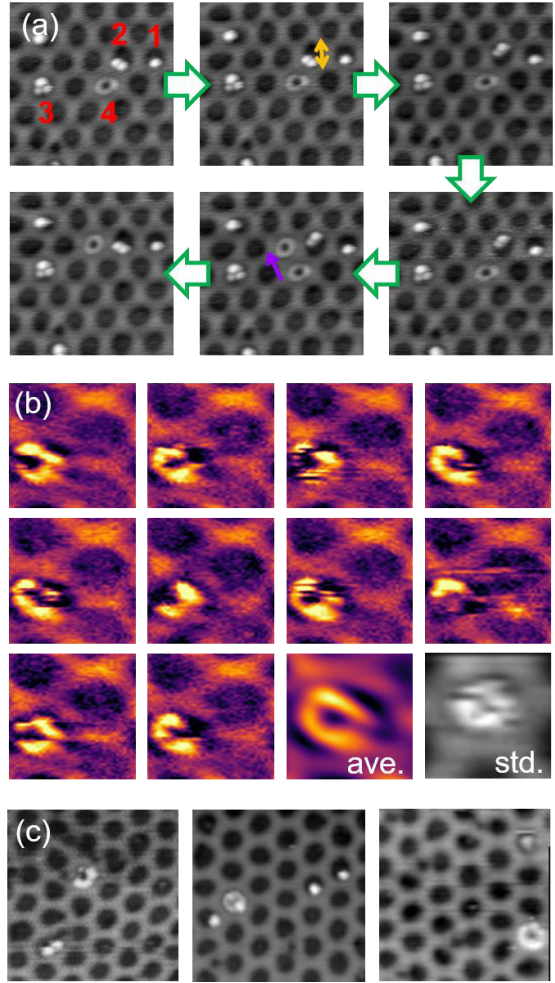

Figure 2. STM images of mobile Pd atoms on h-BN/Rh(111) at LT. (a) A time series measurement shows Pd atoms exhibiting four different shapes: (1) single, (2) double, (3) triple protrusions and (4) rings. Species (2) and (4) move between frames, indicated by the yellow and purple arrows, respectively. (b) Consecutive frames from a FastSTM movie show that a ring appears due to a species running along the pore rim. Averaging 1272 frames from the drift-corrected movie gives a picture of the residence time in different parts of the pore (marked 'ave'). A similar picture is obtained by standard STM (marked 'std.'). (c) STM images in defective h-BN regions show partial rings and rings with uneven brightness. *Imaging parameters:* $V_b$ = -0.6 mV, (a) $I_t$ = 1 pA, 20 x 20 nm$^2$, 2.3 fpm, (b) $I_t$ = 150 pA, 16 fps, approx. 5 x 5 nm$^2$, (c) $I_t$ = 1 pA, 50 pA and 50 pA, respectively, 20 x 20 nm$^2$.

Non-uniform contrast is also often observed in defective h-BN areas. The brighter bottom right corner of the ring in Figure 2c (center) is due to a longer mean residence time in this location compared to the remainder of the pore, and a brightness modulation of the ring in Figure 2c (right) implies several preferred sites. Pd diffusing in a nanomesh pore thus gives us a detailed picture of the potential energy landscape it experiences, showing insurmountable barriers (manifest as partial rings) and differences in residence times (brightness changes). Additionally, the time series measurement in Figure 2a shows that Pd rings can occasionally jump between neighboring pores – in contrast to the single protrusion species which do not diffuse on the time scale of our measurements.

From all of the above observations, we can deduce the relative magnitude of energy barriers for different processes: There is no significant barrier to diffusion along the rim. A wire then constitutes the lowest barrier to diffusion, followed by the barrier to becoming immobilized on a defect. Even more energy is needed for dimer formation, and finally the most difficult process is the detachment of an atom from a defect.

### c) Cluster diffusion in nanomesh pores

We now move on to studying the dynamics of *clusters*. It is inherently difficult to observe cluster dynamics microscopically since non-equilibrium dynamics are completed before we begin a measurement. For example, we usually see stable structural isomers which switch only very rarely,[10] even though the cluster most certainly had to restructure from completely different gas phase shapes due to support interactions.[38] Even steady state diffusion is often stopped prematurely when clusters bind to defects in the substrate. Indeed, we did not observe significant diffusion of as-deposited clusters of any size and instead saw static structures such as those shown in Figure 1. One possibility to activate dynamic processes is thermal activation. Unfortunately, the resulting drift once again means that by the time we can measure stably, the clusters are immobilized on defects in the h-BN film. Due to the smaller footprint, smaller clusters are more mobile than larger ones and supposedly get pinned to defects at lower temperatures. We therefore used the tip to interact with clusters *in situ*, thus triggering cluster dynamics. By shrinking larger clusters, *e.g.* $Pd_{12}$, which have not reached a defect yet, we could generate small mobile clusters on h-BN in single cases, on purpose.

Figure 3 shows a mobile cluster at 375 K which resulted from tip-induced shrinking of $Pd_{12}$. From its height we conclude that it is a single-layer cluster with three to seven atoms, $Pd_{3-7}$. The cluster was highly mobile, "dancing" within the cell for three hours, as shown in Supporting Movie 2. Several different locations are shown in Figure 3a. While the cluster locations seem to be statistically distributed within the pore at a first glance, a careful determination of their apparent center of mass in every frame of the drift corrected FastSTM movie reveals that the cluster strongly prefers six discrete sites in the pore (Figure 3b), corresponding to the six snapshots shown in Figure 3a. The adsorption sites are aligned with the six sides of the nanomesh pore, pointing to the respective neighboring pore, and form an approximately equilateral hexagon. All six sites are occupied with similar probability. From the scatter plot we therefore conclude that these are six energetically equivalent sites, in accordance with the six-fold symmetry of the nanomesh pore. A closer look at the diffusion path of the cluster, shown in Figure 3c, reveals that the cluster moves preferably between adjacent sites. Furthermore, the sites come in pairs, where from any given site, the jump in one direction is more likely than in the other one. Within the pairs of sites, namely 1

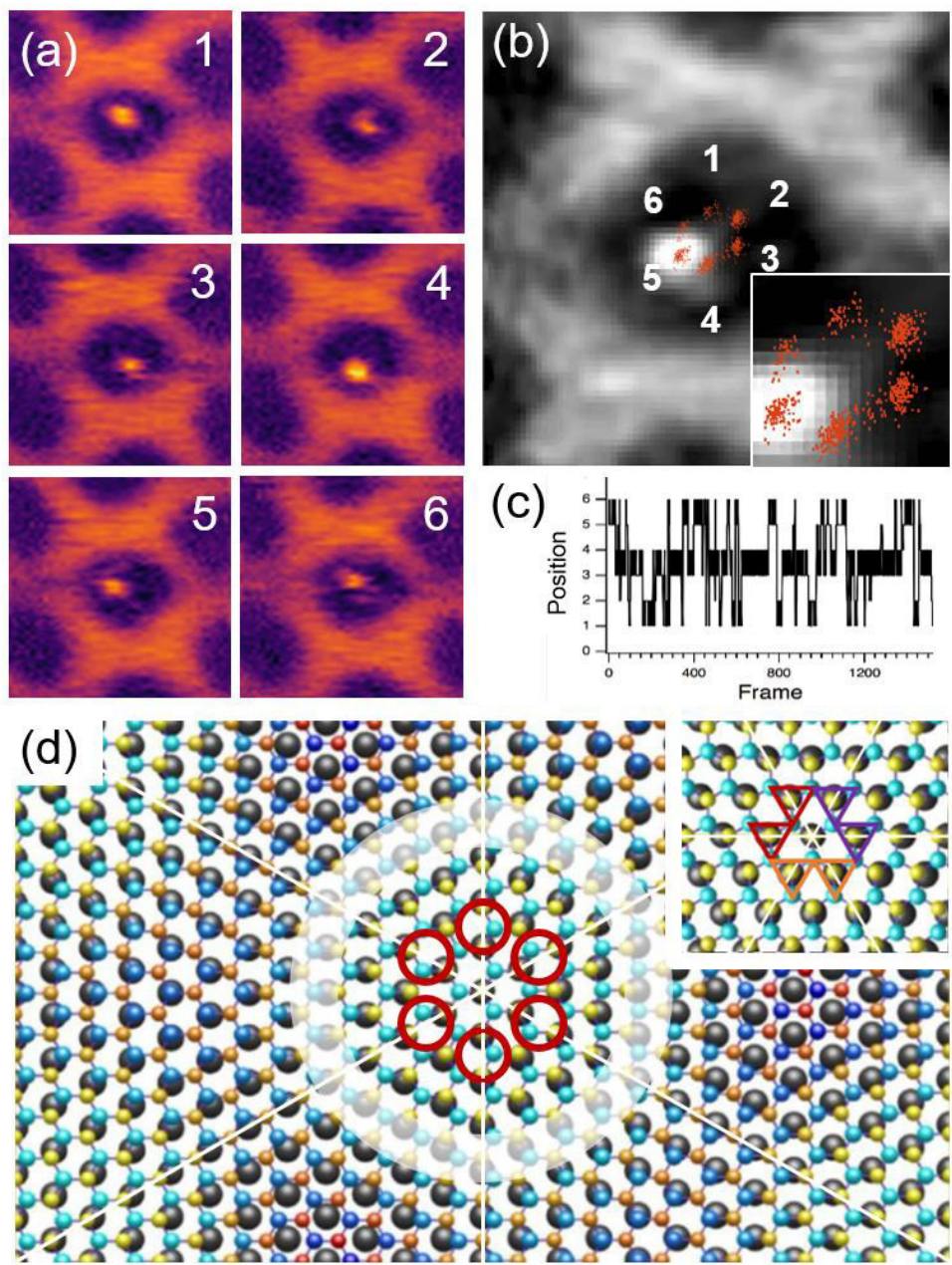

Figure 3. A single-layer Pd cluster dances in an h-BN pore at 375 K. (a) Representative frames of a FastSTM movie show that the cluster moves around the pore. (b) A scatter plot of the cluster positions reveals that it prefers six discrete locations – labeled 1 to 6 in (a) and (b) – which form a hexagon around the center of the pore (magnified in the inset). (c) Extracting the cluster position in every frame shows that from each site the cluster jumps preferentially to one neighboring site

over the other, resulting in three pairs of adsorption sites, *i.e.* the locations marked 1 and 2, 3 and 4, and 5 and 6. (d) Overlaying red circles of the approximate cluster adsorption sites onto a schematic of the h-BN/Rh(111) pore (adapted with permission from Ref. 22. Copyrighted by the American Physical Society; B turquois in the pore and blue on the wire, N yellow in the pore and red on the wire, Rh gray) demonstrates that the maximum brightness is not located centrally over an atom of the boron nitride film. A larger white ring illustrates the diameter of the atom rings shown in Figure 2. The inset illustrates six possible locations of a $Pd_3$ cluster where each atom sits on top of a B atom, at a corner of a triangle. *Imaging parameters:* 16 fps, $V_b$ = -0.6 V, (a) $I_t$ = 200 pA, approx. 6 x 6 $nm^2$, (b) $I_t$ = 300 pA, approx. 5 x 5 $nm^2$.

and 2, 3 and 4, and 5 and 6, the cluster performs frequent jumps. Diffusion between the pairs, *e.g.* from 2 to 3, occurs more rarely. By fitting the probability distribution of residence times with an exponential decay, we obtain a mean residence time of 0.05 s in individual sites (on the order of the frame rate of our measurement) and 0.7 s (*i.e.* 14 times longer) in the pairs of sites. The potential energy landscape thus exhibits a three-fold symmetry due to two different activation barriers connecting the six equivalent sites – again in accordance with the geometric model of the nanomesh.[22]

We changed the tunneling conditions during our FastSTM measurements to rule out tip-induced cluster diffusion. The mean residence time remains largely unchanged for bias voltages between ±1.2 V and setpoint currents from 200 to 1000 pA. By only changing the current while keeping the bias constant, we could further verify that the tip–cluster distance does not influence the steady state dynamics of the cluster. Only when we increase the bias beyond ±1.2 V, the cluster moved around more rapidly and eventually appeared frayed in the STM images. This effect was reversible and the motion returned to the same steady state when the bias voltage was reduced again. The field from a higher voltage thus appears to excite internal vibrational modes of the cluster, which could account for the fuzzy appearance.

With these observations we construct the most plausible model for cluster geometry and location. While Pd atoms diffuse continuously along the pore rim (white ring in Figure 3d), just like other adsorbates,[30–33] the Pd cluster sits more centrally in the pore – but never occupies the center itself. Furthermore, we observe six distinct locations separated by two alternating diffusion barriers. We measure that two neighboring adsorption sites have an average apparent distance of 0.38 nm and are located 0.38 nm from the center of the pore, while the B-B distance of h-BN is 0.25 nm. These locations are marked by red circles in a model of the nanomesh in Figure 3d. The differences between atoms and cluster can be described in the light of several effects: footprint, polarizability and charging. Whereas the footprint and (negative) charging stabilize clusters in the center of the pore, polarizability drives atoms towards the rim of the pore.

Previous studies reported that the atoms of metal clusters and adatoms bind to the B atoms of h-BN in the pores, whereby the interaction is strongest in the center of the pore where the N atoms are located roughly above the Rh support atoms.[37,39,40] Thus, the larger the cluster footprint, the stronger the local interaction with this sticky part of the pore; smaller clusters are increasingly mobile. As a result, we can further narrow down the size of the mobile cluster we observe, since a single-layer $Pd_7$ cluster would have an equal or larger footprint than the original multi-layer, *immobile* $Pd_{12}$ cluster. In addition, to explain the three-fold symmetry of the two diffusion barriers, the cluster footprint has to mirror the same substrate symmetry, $C_{3v}$. As a result, a plausible model for our observation is a $Pd_3$ cluster, as shown in the inset of Figure 3d. It is small enough to be sufficiently mobile, has all atoms on top of a B atom and the center of mass on the high symmetry axes (white lines).

Polarizability and charging effects are also size-dependent. Since the work function of Pd is larger than that of the nanomesh in the center of the pores (4.15 eV),[33] we expect larger clusters with higher electron affinities to have a stronger tendency to be negatively charged than the Pd atom that can bear less charge. As a consequence, the atom's behavior is governed by polarizability. It is attracted to the electrostatic potential gradient of the rim, as described by Dil and coworkers for other polarizable adsorbates.[33] For the cluster, instead, the charge attraction towards the center of the inhomogeneous electrostatic field in the middle of the pore apparently dominates over the polarizability (larger than for atoms) and results in a fairly central location. The reason for avoiding the very pore center might be related to a high inward barrier, similar to that described for analogue adsorption sites of Pd clusters on graphene/Rh(111).[10]

### d) Cluster isomerization

We observed another type of motion after thinning down a cluster with the tip, namely the reversible isomerization. The FastSTM frames in Figure 4 depict such isomerization of a cluster shrunk down from $Pd_{12}$ at 375 K. In Figure 4a, a single-layer cluster switches repeatedly between two apparent locations (see Supporting Movie 3) with a mean residence time of 1.3 s in the site closer to the rim (green) and 3.7 s in the site further away from the rim (white). The two positions of the cluster are approximately one atomic distance apart. Center-of-mass motion of the entire cluster is unlikely because Smoluchowski ripening, *i.e.* the diffusion of entire clusters, has not been observed for $Pd_n$/h-BN/Rh(111) and is energetically unfavorable as many Pd-B bonds would need to be broken and reformed. Instead, a rearrangement of the cluster atoms (peripheral diffusion), resulting in structural isomers with different density of states maxima in STM, is a more plausible scenario. The mean residence times in the two configurations are a factor 2.8 different, which could be a measure of a small energetic difference between the two isomers.

The cluster in Figure 4b switches between two isomers of different apparent brightness. The switching once again occurs repeatedly over many minutes but at a rate 8 times slower. In this case, the cluster resides 2.5 times longer in the bright state closer to the rim than in the darker state, whereas the cluster in Figure 4a preferred to reside further away from the rim. As in all STM measurements, it is important to exclude tip effects as an explanation for the apparent switching behavior. After the initial measurement, we moved the tip to another nearby cluster and recorded a FastSTM movie with identical settings there; the cluster remained static throughout the entire 3-minute movie. Returning to the location of the fused cluster after 5 minutes (bottom right two frames of Figure 4b and Supporting Movie 4), the switching process continued unchanged. We therefore conclude that we observe a real effect and not a measurement artifact.

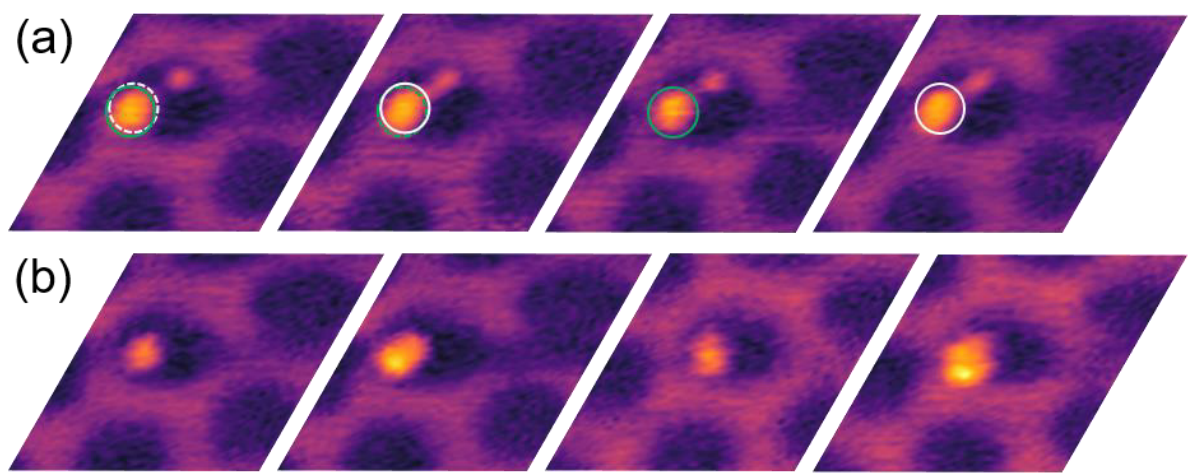

Figure 4. Reversible isomerization of small Pd clusters in an h-BN pore observed in FastSTM at 375 K. The frames were skewed to correct for distortions. (a) A single-layer Pd cluster switches repeatedly between two apparent locations in the nanomesh pore, marked by green and white circles, respectively. Dashed circles provide a direct comparison of both positions. Representative, non-consecutive frames are shown. Note that the small feature in the top of the cell is likely a static defect and seemingly does not participate in the cluster dynamics. (b) A small Pd cluster switches between two isomers of different apparent height. The left two (non-consecutive) frames are taken from the same movie, the right two from another, later movie which returned to this location. *Imaging parameters:* approx. 6 x 6 nm$^2$, 16 fps, $I_t$ = 300 pA, $V_b$ = -0.6 mV.

CONCLUSIONS

The present work demonstrates the versatility of FastSTM when it comes to investigating the dynamics and energetics of metal clusters supported on a thin film. We have shown that the diffusion of Pd atoms and clusters on h-BN/Rh(111) is strongly influenced by defects. At first sight, clusters appear completely immobile within a nanomesh pore. However, shrinking a cluster by removing several atoms with the STM tip increased its mobility significantly. Smaller clusters interact more weakly with the h-BN substrate due to their smaller footprint and can thus overcome barriers to diffusion that larger ones cannot. Pd atoms are much more mobile still, and sinter significantly at RT. Even at 150 K, the spatial distribution of Pd atoms implies that they moved around the surface until they were caught by a defect. In fact, we could observe several highly mobile Pd atoms, diffusing along the rim of a nanomesh pore and even jumping between different pores.

The FastSTM technique allowed us to investigate the diffusing clusters and atoms with high spatial and temporal resolution. While atoms diffuse along the rim continuously, small clusters show a preference for distinct adsorption sites. By extracting the diffusion trace of a single-layer cluster, we revealed its preference for six equivalent adsorption sites located close to the center of the nanomesh pore and two different types of diffusion barriers connecting these sites. These dynamics can be explained as an interplay between size-dependent footprint and charge effects. While the diffusion of clusters with larger footprints is governed by strong local interactions with the center of the pore, polarizable atoms remain captured by the electrostatic field gradient at the rim. Charging of larger clusters might additionally favor adsorption closer to the center of the pores.

**Supporting Information.** (a) Tip induced cluster decay. (b) Distribution of Pd atoms across nanomesh pores. (c) Dynamics of Pd atoms and clusters in nanomesh pores. (d) Characterization of boron nitride defect density by TPD.

ACKNOWLEDGEMENTS

The authors would like to thank Willi Auwärter and Sebastian Günther for useful discussions. This work was supported by the Deutsche Forschungsgemeinschaft (Research grants ES 349/1-2 and HE 3454/18-2) and the Project "Nanoscience Foundries and Fine Analysis", NFFA (EU Programme Horizon 2020). BAJL gratefully acknowledges a Research Fellowship from the Alexander von Humboldt Foundation and a Marie Skłodowska-Curie Individual Fellowship under grant ClusterDynamics (no. 703972) from the European Union's Horizon 2020 research and innovation program.

ORCID

| | |
|---|---|
| Barbara A. J. Lechner: | 0000-0001-9974-1738 |
| Friedrich Esch: | 0000-0001-7793-3341 |

TOC GRAPHIC

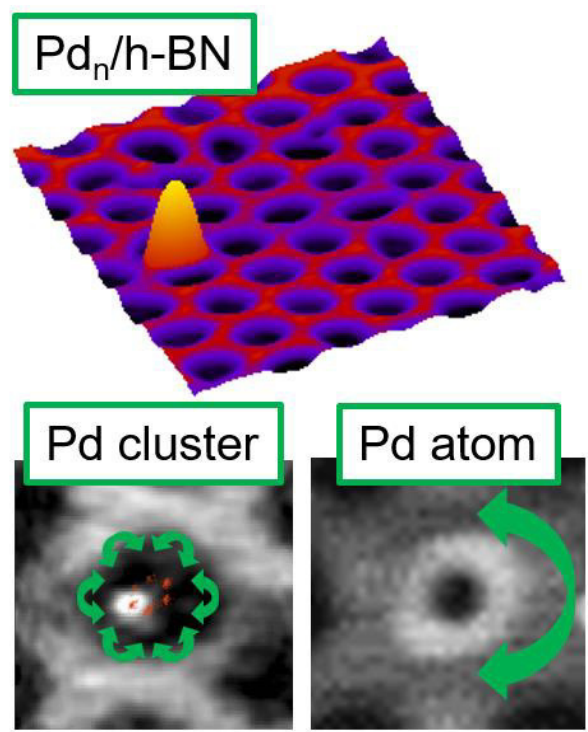

Supporting Information

for

A Microscopy Approach to Investigating the Energetics of Small Supported Metal Clusters

*Barbara A. J. Lechner, Fabian Knoller, Alexander Bourgund, Ueli Heiz, and Friedrich Esch\**

Chair of Physical Chemistry, Department of Chemistry & Catalysis Research Center, Technical University of Munich, Lichtenbergstr. 4, 85748 Garching, Germany

\* friedrich.esch@tum.de

### a) Tip-induced cluster decay

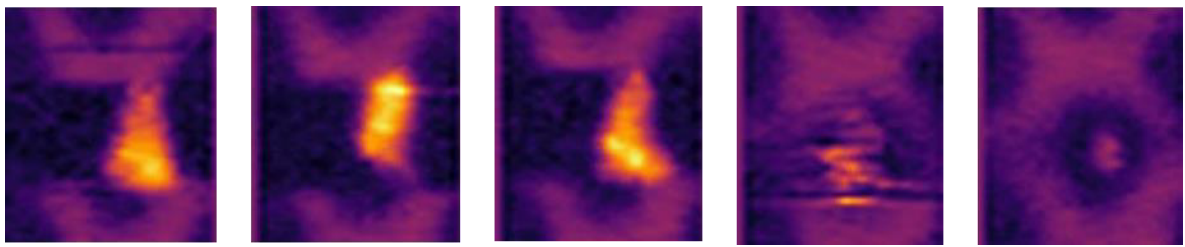

Figure S1. After slowing down the feedback, several atoms of a Pd$_{12}$ cluster are removed by the tip during the first six frames of a FastSTM measurement, leaving behind a single-layer cluster. *Imaging parameters:* 16 fps, approximately 4 x 5 nm$^2$, $I_t$ = 150 pA, $V_b$ = -0.6 mV.

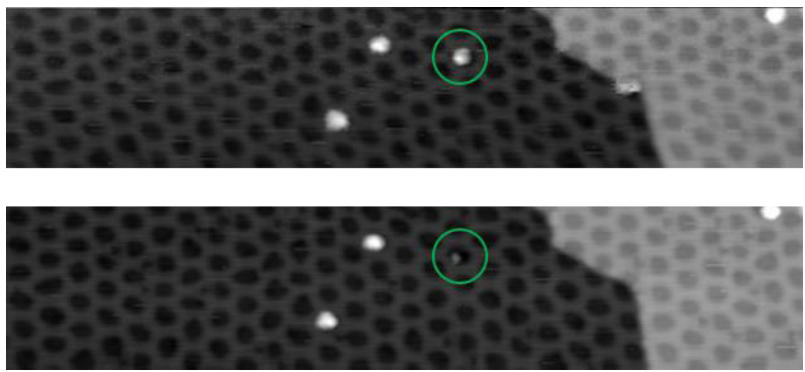

Figure S2. STM measurements at 375 K recorded before (top) and after (bottom) the FastSTM movie presented in Figure S1. The pore in question is circled in green. As the missing atoms could not be accounted for in the vicinity of the cluster, they were likely picked up by the tip. *Imaging parameters:* 100 x 20 nm$^2$, $I_t$ = 15 pA, $V_b$ = -0.6 mV.

b) **Distribution of Pd atoms across nanomesh pores**

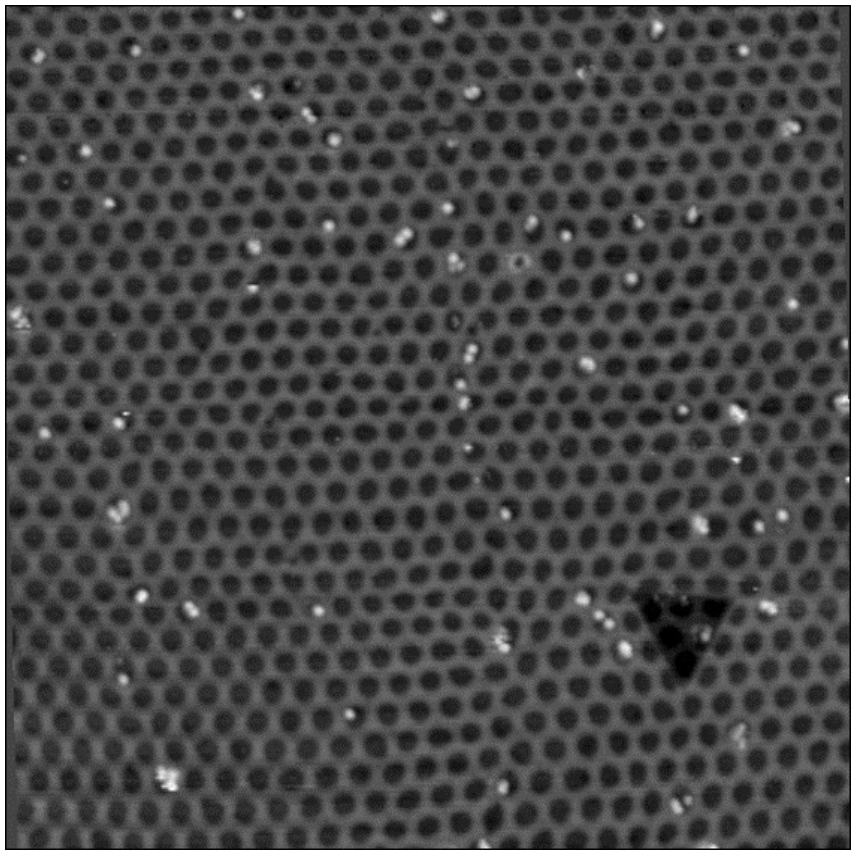

Figure S3. A 100 x 100 nm² STM image of Pd$_1$/h-BN/Rh(111) at LT shows a significant number of seemingly doubly and triply occupied nanomesh pores as well as local clustering of several atoms around some defective areas of the h-BN film, implying that atoms are pinned at defects. *Imaging parameters:* $I_t$ = 1 pA, $V_b$ = -0.6 mV.

c) **Dynamics of Pd atoms and clusters in nanomesh pores**

Supporting Movie 1. FastSTM movie of a Pd atom diffusing in a nanomesh pore which appeared as a ring in standard STM measurements. While the time resolution is still not sufficient to show the atom at a single location in each frame, the ring appears shorter and moves between the frames of the movie, demonstrating that we are imaging a mobile adatom. *Imaging parameters:* 16 fps, approximately 5 x 5 nm$^2$, $I_t$ = 150 pA, $V_b$ = -0.6 mV.

Supporting Movie 2. FastSTM movie of a Pd cluster diffusing between six specific sites in a nanomesh pore. *Imaging parameters:* 16 fps, approximately 6 x 6 nm$^2$, $I_t$ = 200 pA, $V_b$ = -0.6 mV.

Supporting Movie 3. FastSTM movie of a Pd cluster switching reversibly between two apparent locations. *Imaging parameters:* 16 fps, approximately 6 x 6 nm$^2$, $I_t$ = 300 pA, $V_b$ = -0.6 mV.

Supporting Movie 4. FastSTM movie of a Pd cluster switching reversibly between two isomers of different brightness. *Imaging parameters:* 16 fps, approximately 6 x 6 nm$^2$, $I_t$ = 300 pA, $V_b$ = -0.6 mV.

All movies show the elapsed real time in the top right corner and the frame index in the top left corner, where "u" stands for a frame measured upwards and "d" downwards.

### d) Characterization of boron nitride defect density by TPR

Temperature programmed reaction measurements of the $O_2$+CO reaction were used to estimate the order of magnitude of the defect density of our boron nitride films on Rh(111). We calibrated the sensitivity with a saturation coverage of $O_2$ and CO on a preparation of 2 % $Pd_{\sim19}$ clusters on h-BN/Rh(111), assuming a saturation coverage of approx. 0.5 CO molecules per Pd atom. Comparing the $CO_2$ desorption peak off a pristine, bare h-BN surface, we find that the film has an average defect density of up to 0.8% on average, comparable to the values obtained in ref. 1.

(1) Mertens, S. F. L. Copper Underpotential Deposition on Boron Nitride Nanomesh. *Electrochim. Acta* **2017**, *246*, 730–736.